\documentclass[useAMS,usenatbib,pdflscape]{mn2e}
\usepackage{graphicx}
\usepackage{epstopdf}
\usepackage{amsmath}
\usepackage{amssymb}
\usepackage{bm}
\usepackage{colortbl,dcolumn}

\usepackage{rotating}
\usepackage{pdflscape}
\usepackage{appendix}
\title[Large-scale Winds Dynamics]{Large-scale Dynamics of Winds Driven by Line Force from a Thin Accretion Disk}
\author[Zhu, Bu, Yang, Yuan \& Lin]{Yi Zhu$^1$ , De-Fu Bu$^2$\thanks{Corresponding author: dfbu@shao.ac.cn} , Xiao-Hong Yang$^3$\thanks{Corresponding author: yangxh@cqu.edu.cn} , Feng Yuan$^2$ and Wen-Bin Lin$^{1,4}$ \\
$^{1}$School of Nuclear Science and Technology, University of South China, Hengyang 421001, China \\
$^{2}$Shanghai Astronomical Observatory, Chinese Academy of Sciences, 80 Nandan Road, Shanghai 200030, China \\
$^{3}$Department of Physics, Chongqing University, Chongqing 400044, China \\
$^{4}$School of Physical Science and Technology, Southwest Jiaotong University, Chengdu 610031, China}
\begin{document}


\maketitle


%
\begin{abstract}
Winds play a significant role in active galactic nuclei feedback process. Previous simulations studying winds only focus on a small dynamical range. Therefore, it is unknown how far the winds can go and what the properties of the winds will be if they can move to large radii. We perform simulations to study the large scale dynamics of winds driven by line force. We find that the properties of the winds depend on both black hole mass ($M_{BH}$) and accretion disk luminosity. When the accretion disk luminosity is $0.6L_{edd}$ ($L_{edd}$ being Eddington luminosity), independent of $M_{BH}$, the winds have kinetic energy flux exceeding $1\% L_{edd}$ and can escape from the black hole potential. For the case with the accretion disk luminosity equaling 0.3$L_{edd}$, the strength of the winds decreases with the decrease of $M_{BH}$. If $M_{BH}$ decreases from $10^9$ to $10^6$ solar mass ($M_\odot$), the winds kinetic energy flux decreases from $\sim 0.01 L_{edd}$ to $ \sim 10^{-6} L_{edd}$. In case of $M_{BH}\geq 10^7 M_\odot$, winds can escape from black hole potential. In the case of $M_{BH}=10^6 M_\odot$, the winds can not escape. We find that for the ultra-fast winds observed in hard X-ray bands (\citealt{Gofford et al. 2015}), the observed dependence of the mass flux and the kinetic energy flux on accretion disk luminosity can be well produced by line force driven winds model. We also find that the properties of the ultra-fast winds observed in soft X-ray bands can be explained by the line force driven winds model.
\end{abstract}


\section{Introduction}
There are two main accretion models, namely hot accretion flow and standard thin disk. Hot accretion flow is the accretion model for low-luminosity systems. The low-luminosity systems include low-luminosity AGNs and the hard and quiescent states of black hole X-ray binaries (see \citealt{Yuan and Narayan 2014} for a review). The standard thin disk (\citealt{Shakura et al. 1973}) powers luminous systems such as luminous AGNs and the soft state of black hole X-ray binaries.

There are more and more observational evidences for the presence of winds from both hot accretion flows and cold thin disks. Indirect observational evidences of winds from hot accretion flows have been reported for low-luminosity AGNs (\citealt{Crenshaw and Kraemer 2012}; \citealt{Cheung et al. 2016}; \citealt{Wang et al. 2013}; \citealt{Almeida et al. 2018}; \citealt{Ma et al. 2019}; \citealt{Park et al. 2019}). Recently, winds from hot accretion flows have been directly observed in the low-luminosity AGN M81 (\citealt{Shi et al. 2021}). Winds from hot accretion flows are also detected in black hole X-ray binaries (\citealt{Homan et al. 2016}; \citealt{Munoz-Darias et al. 2019}). Winds from cold thin disks are frequently observed for both luminous AGNs (e.g., \citealt{Crenshaw et al. 2003}; \citealt{Tombesi et al. 2010,Tombesi et al. 2014}; \citealt{Liu et al. 2013}; \citealt{Gofford et al. 2015}; \citealt{King and Pounds 2015}; \citealt{He et al. 2019}) and the soft state of black hole X-ray binaries (\citealt{Neilsen and Homan 2012}; \citealt{D'iaz Trigo et al. 2016}; \citealt{Homan et al. 2016}; \citealt{You et al. 2016}).

The production mechanisms and properties of winds from hot accretion flows have been intensively investigated (e.g., \citealt{Yuan et al. 2012,Yuan et al. 2015}; \citealt{Narayan et al. 2012}; \citealt{Bu and Gan 2018}). It is found that winds are launched and accelerated by the combination of gas pressure gradient, magnetic pressure gradient and centrifugal forces.

For a cold thin disk, there are three main mechanisms to produce winds. They are thermal (\citealt{Begelman 1983}; \citealt{Woods et al. 1996}), magnetic (\citealt{Blandford and Payne 1982}) and radiation line force (\citealt{Murray et al. 1995}) mechanisms. The radiation line force mechanism has been studied by numerical simulations (\citealt{Proga et al. 2000}; \citealt{Nomura et al. 2016}; \citealt{Nomura and Ohsuga 2017}; \citealt{Nomura et al. 2020}). It is found that line force can effectively drive high speed winds from the inner region (inside 100 Schwarzschild radius) of a cold thin accretion disk. The ultra-fast outflows that found in luminous AGNs (\citealt{Tombesi et al. 2010}) are believed to be driven by line force (\citealt{Nomura et al. 2016}; \citealt{Nomura and Ohsuga 2017}; \citealt{Yang et al. 2021a}; \citealt{Yang 2021b}). \citet{Zhu and Stone 2018} use numerical simulations to investigate the magnetic driven winds mechanism from a thin disk. It is found that very weak winds can be magnetically driven from a thin disk. \citet{Yang et al. 2021a} and \citet{Yang 2021b} investigate winds driven simultaneously by both line force and magnetic field. They find that with the including of magnetic field, the opening angle of the winds becomes larger. They also find that the ultra-fast outflows in radio-lond AGNs may be driven by the combination of line and Lorentz forces.

All the above mentioned simulations studying both hot accretion flows and cold thin disks have a very limited computational domain. The outer boundary is usually located at several hundreds or thousands Schwarzschild radius. When winds generated by the accretion flows/disks cross the outer radial boundary to large radii, the dynamics of winds can not be answered by the simulations. For example, how far can winds arrive? How does the kinetic power of winds evolve? These questions are important for AGNs feedback study. It is widely believed that winds play a significant role in the interaction between the AGNs and their host galaxies (e.g., \citealt{Ciotti et al. 2010,Ciotti et al. 2017}; \citealt{Ostriker et al. 2010}; \citealt{Weinberger et al. 2017}; \citealt{Yuan et al. 2018}). Winds can very effectively affect the properties of the gas surrounding the AGNs (\citealt{Yuan et al. 2018}). In order to study the winds feedback, we have to know the properties of the winds at large scale.

\citet[2020b]{Cui et al. 2020a} study the propagation of winds at large scale by analytical and simulation methods. They investigate winds both from hot accretion flows and cold thin disks. In their works, the inner radial boundary is located at 3000 Schwarzschild radius. They inject winds from the inner boundary. The properties of the injected winds generated by hot accretion flows are set according to the simulation results of \citet{Yuan et al. 2015}. The properties of the injected winds generated by a thin disk are taken from observations (\citealt{Gofford et al. 2015}). There are several assumptions in their work. First, radiation pressure is neglected. This assumption is suitable for winds from hot accretion flows because radiation of hot accretion flows is weak. However, for winds from cold thin disks, this assumption is problematic. As mentioned above, radiation line force plays an important role in driving winds process. When winds propagate at large-scale, line force may continuously accelerate winds. \citet{Cui et al. 2020a} show that when winds from cold thin disks have negative Bernoulli parameter, it will just propagate a very short distance from the injected inner boundary of their simulations. This result may be due to the neglect of radiation line force. If radiation line force is included, winds from cold thin disks may propagate to very large radii even they initially have a negative Bernoulli parameter. Second, the radiative cooling is neglected. This assumption is suitable for winds from hot accretion flows because of the low radiative efficiency of hot accretion flows. However, winds from cold thin disks have much higher density and correspondingly strong radiative cooling, the neglect of radiative cooling may be problematic. It is very necessary to study the large-scale dynamics of winds from cold thin disks.

In this paper, we perform numerical simulations to study the large scale dynamics of winds driven by line force. In our simulations, both radiation pressure force and radiative cooling/heating are included. We first simulate the radiation line force driven winds in a dynamical range of $30-1500 r_s$. After a quasi-steady state is formed, we time-average the properties of winds. Then, we perform simulations in a dynamic range of $1500-10^6 r_s$. Winds are injected into the computational domain at $1500 r_s$. The properties of the injected winds are from the time-averaged value of winds at $1500r_s$ from the small scale simulations.

The paper is structured as follows. In section 2, we describe our models and methods; In section 3, the results are presented; We discuss the limitations of this work in Section 4. A summary is presented in Section 5.

\section{Model And Method }

\subsection{Basic Equations}
We employ spherical coordinates ($r,\theta,\phi$) and use PLUTO code (\citealt{Mignone et al. 2007,Mignone et al. 2012}; \citealt{Yang et al. 2021a}) to solve the hydrodynamics equations as follows (Proga et al. 2000)
\begin{equation}
 \frac{d\rho}{d t} + \rho\nabla\cdot {\bm \upsilon} = 0,
\end{equation}
\begin{equation}
\rho\frac{d {\bm \upsilon}}{dt}=-\nabla P-\rho \nabla\psi+\rho {\bm F}_{rad},
\end{equation}
\begin{equation}
\rho\frac{d}{dt}\Big(\frac{e}{\rho}\Big)=-P\nabla\cdot {\bm \upsilon}+\rho\zeta.
\end{equation}
Here, $\rho$, $\bm\upsilon$, $P$, $e$, and $\psi$ are density, velocity, gas pressure, internal energy, and gravitational potential, respectively. $\bm F_{rad}$ is the radiation pressure force on unit mass including Compton scattering and line force. $\zeta$ denotes the net cooling/heating rate. We employ an adiabatic equation of state $P=(\gamma-1)e$ with $\gamma=5/3$. We apply the pseudo-Newtonian potential, $\psi=-GM_{BH}/(r_c-r_{s})$ (\citealt{Paczy'nsky and Wiita 1980}), where $M_{BH}$, $G$, $r_{s}$, and $r_c$ are the BH mass, the gravitational constant, the Schwarzschild radius, and the distance from a point to the black hole, respectively. The calculation of net cooling/heating rate $\zeta$ is referred to \citet{Proga et al. 2000}.

\begin{figure*}
\begin{center}
\includegraphics[width=0.45\textwidth]{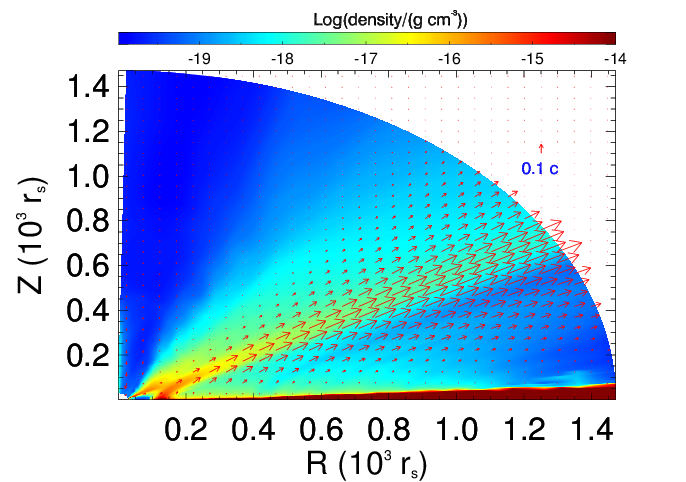}
\includegraphics[width=0.45\textwidth]{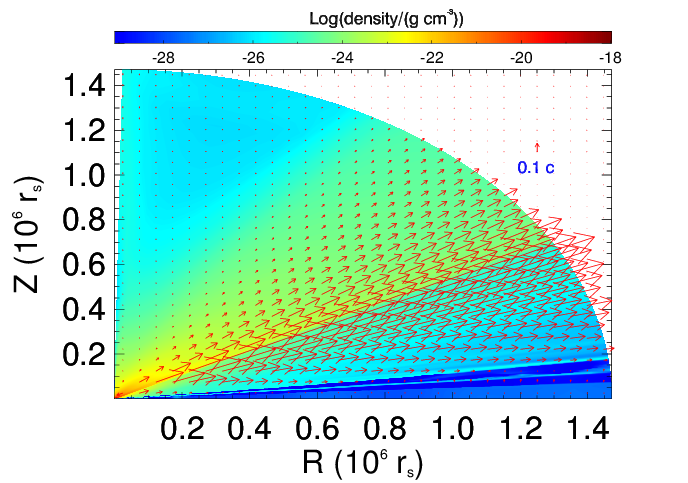}
 \caption{Time-averaged density (color) and velocity (arrows) for model RHD86. Left panel is for small scale simulations and right panel is for large scale simulations. \label{Fig:dv86}}
\end{center}
\end{figure*}
\subsection{Model Setup}
AGNs are powered by accretion disk around a central black hole. We assume that the accretion disk mainly emits UV photons. There is also a very compact hot corona above and below the central black hole (e.g., \citealt{Uttley et al. 2014}). The hot corona radiates X-rays. In this paper, we assume that the hot corona is spherically distributed around the central black hole. The X-rays from the corona can irradiate the accretion disk. The radiation intensity from the surface of the accretion disk is given by \citet{Proga et al. 1998}
\begin{equation}
\begin{aligned}
&&I_{D}(r_{D})=\frac{3GM_{BH}\dot{M_{a}}}{8\pi^{2}r_{\ast}^{3}}\bigg(\frac{r_{\ast}^{3}}{r_{D}^{3}}\bigg[1-\bigg(\frac{r_{\ast}}{r_{D}}\bigg)^{1/2}\bigg]+\\
&&\frac{f_x}{3\pi}\bigg\{\arcsin\frac{r_{\ast}}{r_{D}}-\frac{r_{\ast}}{r_{D}}\bigg[1-\bigg(\frac{r_{\ast}}{r_{D}}\bigg)^{2}\bigg]^{1/2}\bigg\}\bigg),
\end{aligned}
\end{equation}
where $\dot{M}_{a}$ is the mass accretion rate of the accretion disk, $r_{\ast}$ is the inner edge of the disk and $r_{\ast}=3 r_{s}$, $r_{D}$ is the radial position on the disk surface measured from the central black hole. $f_x$ denotes the radio of the X-ray luminosity from the corona to the disk luminosity. The effective temperature is $T_{eff}=(\pi I_{D}(r_{D})/\sigma)^{1/4}$, with $\sigma$ being the stefan-Boltzmann constant. When we calculate the disc radiation using Equation (4), we only consider the region which has a effective temperature higher than 3000K contributing to the line force.

The total disk luminosity is $L_{D}=\varepsilon L_{\rm edd}=GM_{BH}\dot M_{a}/6r_s$, with $\varepsilon$ being the Eddington ratio. Observationally, there is a relationship between the X-ray and UV emissions in AGNs (\citealt{Lusso and Risaliti 2016}). In general, more luminous AGNs will have fainter X-rays. We refer to \citet{Yang et al. 2021a} for the detailed calculation of the X-ray luminosity based on the disk luminosity.

The gas above the thin disk is irradiated by both the spherically distributed X-ray photons from the corona and the UV photons from the accretion disk surface. We put the surface of the accretion disk on the plane of $\theta=\pi/2$. The black hole is one disk scale-height below the plane of $\theta=\pi/2$. For the radiation pressure dominated inner region of the accretion disk, the scale height does not vary much with radius. Therefore, for simplicity, we assume that the disk scale height is a constant of radius. The scale height of the disk is $3.1\varepsilon r_s$ (\citealt{Nomura and Ohsuga 2017}). In this paper, we also adopt this value. Thus, the black hole is located at $r=3.1\varepsilon r_s, \theta=\pi$.

The calculation of the radiation force is referred to appendix in \citet{Proga et al. 1998} and the multiplier (\textbf{M}) of line force is referred to equation (11)-(16) in \citet{Proga et al. 2000}.

\subsection{Computational Domain and Boundary Conditions}
We first simulate the line force driven winds for various black hole masses and luminosities in a radial range from $30r_{s}$ to $1500r_{s}$. In these small scale simulations, we set the accretion disk properties at the midplane and the initial gas above the midplane as follows. The density of the surface of the accretion disk is set according to the standard thin disks model (\citealt{Nomura and Ohsuga 2017})
\begin{equation}
\rho(\theta=\pi/2)=\left\{
\begin{array}{lll}
5.24\times10^{-4}(\frac{M_{BH}}{M_{\odot}})^{-1}(\frac{\varepsilon}{\eta})^{-2}(\frac{r}{r_{s}})^{3/2} {\rm g \ cm^{-3}}~,\\
 ~~~~\text{for}~{r\leq 18(\frac{M_{BH}}{M_{\odot}})^{2/21}(\frac{\varepsilon}{\eta})^{16/21}r_{s}}\\
4.66(\frac{M_{BH}}{M_{\odot}})^{-7/10}(\frac{\varepsilon}{\eta})^{2/5}(\frac{r}{r_{s}})^{-33/20}{\rm g \ cm^{-3}}~,\\
~~~~\text{for}~{r> 18(\frac{M_{BH}}{M_{\odot}})^{2/21}(\frac{\varepsilon}{\eta})^{16/21}r_{s}}
\end {array} \right.
\end{equation}
with $M_\odot$ and $\eta=0.0833$ being solar mass and energy conversion rate, respectively. The temperature of the disk surface is set to the effective temperature. We assume that the initial gas above the thin disk is locally isothermal with $T(r,\theta)=T_{eff}(r\sin\theta)$. We assume that initially the gas above the thin disk is in hydrostatic equilibrium. Therefore, its density can be written as
\begin{equation}
\begin{aligned}
\rho(r,\theta)=\rho(\theta=\pi/2)\exp(-\frac{GM_{BH}}{2c_{s}^{2}r\tan^{2}\theta}).
\end{aligned}
\end{equation}
We assume that initially the poloidal velocity of the gas above the disk is zero ($\upsilon_r=\upsilon_\theta=0$). The rotational velocity is set as $\upsilon_\phi (r, \theta)=({GM_{BH}/r})^{1/2}sin(\theta)r/(r-r_s)$ to meet the equilibrium between the black hole gravity and the centrifugal force. For these small scale simulations, at both the inner and outer radial boundary, we employ outflow boundary conditions. Axis-of-symmetry boundary conditions are applied at the pole $\theta=0^{\circ}$.

After the quasi-steady state is achieved, we time-average the simulation data to get the properties of gas (including both winds and inflow) at $r=1500r_{s}$.

For the large scale simulations, our computational domain is from $1.5\times10^{3}r_{s}$ to $1.5\times10^{6}r_{s}$. At the inner boundary ($1500r_s$), the physical variables (density, velocity, internal energy) are set equaling to the time-averaged values from the small scale simulations at $1500r_s$. For such settings, winds are injected into the computational domain from the inner radial boundary. For the outer radial boundary, we employ outflow boundary conditions. We employ axis-of-symmetry and equator-of-symmetry boundary conditions at the pole $\theta=0^{\circ}$ and the equatorial plane $\theta=90^{\circ}$, respectively. Note that we do not put a accretion disk at the midplane in these large scale simulations. Initially, low density gas is put in the computational domain with $\rho_{0}(r)=2.25\times10^{-14}r_{s}^{2}/r^{2}$. The initial gas density is too low to affect the dynamics of the winds injected from the inner radial boundary. The initial gas has zero velocity.

We employ non-uniform grids to discretize the computational domain. For the small scale simulations, our resolution is $144\times160$. For the large scale simulations, our resolution is  $300\times160$. In the radial direction, we use non-uniform grids to resolve the inner region. We set $(\triangle r)_{i+1}/(\triangle r)_{i} =1.04 $. For the grids in the $\theta$ direction, we set the grids as follows. There are the 16 zones uniformly distributed over the angular range of $0^{\circ}-15^{\circ}$ while the 144 zones are non-uniformly distributed over the angular range of $15^{\circ}-90^{\circ}$ in the $\theta$ direction. The angular size ratio is $(\triangle \theta)_{i+1}/(\triangle \theta)_{i} =0.970072$ to achieve good spatial resolution near the disk surface.

\subsection{Calculation of the Optical Depth}
Before defining the optical depth, we first define the ionization parameter of gas
\begin{equation}
\xi=\frac{4\pi F_{X}}{n},
\end{equation}
where $F_{X}$ is the X-ray flux and $n$ is the number density of the gas.

For the small scale simulations, we calculate the optical depth for the X-ray $\tau_{X}(r,\theta)$ from the corona and the optical depth for the UV radiation $\tau_{UV}(r,\theta)$ from the accretion disk as follows
\begin{equation}
\tau_{X}(r,\theta)=\int_{30 r_{s}}^{r} \sigma_{X}\rho(r',\theta)dr',
\end{equation}
\begin{equation}
\tau_{UV}(r,\theta)=\int_{30 r_{s}}^{r} \sigma_{e}\rho(r',\theta)dr'.
\end{equation}

In the large scale simulations, both the X-ray and UV photons are from small scale. Therefore, in order to calculate the optical depths in large scale simulations, we first need to calculate the optical depths across the entire domain of the small scale simulations. They are calculated as
\begin{equation}
\tau_{0,X}(\theta)=\int_{30 r_{s}}^{1500r_s} \sigma_{X}\overline{\rho}(r',\theta)dr',
\end{equation}
\begin{equation}
\tau_{0,UV}(\theta)=\int_{30 r_{s}}^{1500r_s} \sigma_{e}\overline{\rho}(r',\theta)dr',
\end{equation}
$\overline{\rho}$ is the time-averaged density for the small scale simulations.

Thus, the optical depths for the X-ray and UV photons in the large scale simulations are written as
\begin{equation}
\tau_{X}(r,\theta)=\tau_{0,X}(\theta)+\int_{1500 r_{s}}^{r} \sigma_{X}\rho(r',\theta)dr',
\end{equation}
\begin{equation}
\tau_{UV}(r,\theta)=\tau_{0,UV}(\theta)+\int_{1500 r_{s}}^{r} \sigma_{e}\rho(r',\theta)dr'.
\end{equation}

In the above equations, $\sigma_{e}$ is the mass-scattering coefficient for free electrons and is set to be $0.4 \text{g}^{-1} \text{cm}^{2}$. As same as \citet{Proga et al. 2000}, we set $\sigma_{X}=\sigma_{e}$ for $\xi\geq10^{5}\text{erg cm s}^{-1}$ and $\sigma_{X}=100\sigma_{e}$ for $\xi<10^{5}\text{erg cm s}^{-1}$ in order to include the effects of photoelectronic absorption.

\begin{table*} \caption{Summary of Models }
\setlength{\tabcolsep}{2mm}{
\begin{tabular}{cccccccccc}
\hline \hline
 Cold disk winds   &  $f_x$   &  $\varepsilon$ & $M_{BH}$     & $\dot{M}_{W}$   & $\dot{P}_{W}$ & $\dot{E}_{W}$ & $\dot{M}_{W}'$   & $\dot{P}_{W}'$ & $\dot{E}_{W}'$ \\

                   &           &                & ($M_{\odot}$) & ($10L_{edd}/c^{2}$)  & ($L_{edd}/c$) & ($L_{edd}$)& ($10L_{edd}/c^{2}$)  & ($L_{edd}/c$) & ($L_{edd}$) \\
(1)                & (2)       &  (3)           &     (4)       &        (5)      &        (6)       &        (7)   &        (8)      &        (9)       &        (10)         \\

\hline\noalign{\smallskip}
RHD86  & $6.01\times10^{-2}$ & 0.6 & $10^{8}$ & $0.128$ & $0.208$ & $0.0374$& $0.127$ & $0.262$ & $0.0637$ \\
RHD96  & $4.26\times10^{-2}$ & 0.6 & $10^{9}$ & $0.100$ & $0.101$ & $0.0112$& $0.101$ & $0.112$ & $0.0138$\\
RHD76  & $8.48\times10^{-2}$ & 0.6 & $10^{7}$ & $0.137$ & $0.188$ & $0.0289$& $0.135$ & $0.248$ & $0.0531$\\
RHD66  & $1.20\times10^{-1}$ & 0.6 & $10^{6}$ & $0.104$ & $0.0939$ & $0.00962$& $0.104$ & $0.126$ & $0.0200$\\
RHD83  & $8.96\times10^{-2}$ & 0.3 & $10^{8}$ & $0.0681$ & $0.0889$ & $0.0130$& $0.0675$ & $0.0980$ & $0.0165$\\
RHD93  & $6.35\times10^{-2}$ & 0.3 & $10^{9}$ & $0.0629$ & $0.0665$ & $0.00791$& $0.0666$ & $0.0743$ & $0.00936$\\
RHD73  & $1.26\times10^{-1}$ & 0.3 & $10^{7}$ & $0.0163$ & $0.0179$ & $0.00228$& $0.0156$ & $0.0178$ & $0.00232$\\
RHD63  & $1.78\times10^{-1}$ & 0.3 & $10^{6}$ & $1.56\times10^{-3}$ & $1.35\times10^{-4}$ & $2.91\times10^{-6}$& $1.75\times10^{-4}$ & $5.53\times10^{-5}$ & $1.81\times10^{-6}$\\



\hline\noalign{\smallskip}
\end{tabular}}
\begin{flushleft}
 Columen 1: model names; column 2: the ratio of the X-ray luminosity to the disk luminosity; column 3 ; the Eddington ratio $(\varepsilon=L_{D}/L_{edd})$ of disk luminosity; column 4 : the mass of the black hole; column 5-7 are the time-averaged values of winds measured at the outer radial boundary ($r=1500r_{s}$) of the small scale simulations; column 8-10 are the time-averaged values of winds measured at the outer radial boundary ($r=1.5\times10^{6}r_{s}$) of the large scale simulations.
\end{flushleft}

\end{table*}

\section{Results}

\begin{figure}
\begin{center}
\includegraphics[width=0.5\textwidth]{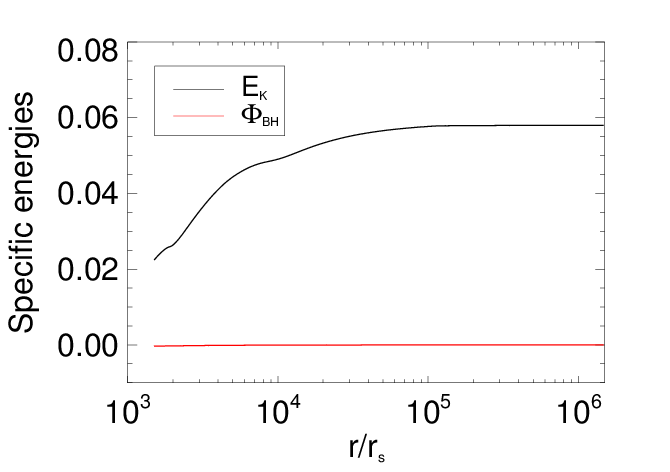}
\caption{Time-averaged specific kinetic energy ($1/2(\upsilon/c)^2$, black line) and gravitational energy ($GM_{BH}/(r-r_s)c^2$, red line) for model RHD86. \label{Fig:Bernoulli}}
\end{center}
\end{figure}
The results of the small scale simulations in the range of $30-1500r_s$ have been discussed by \citet{Nomura et al. 2016}, \citet{Nomura and Ohsuga 2017}, \citet{Yang et al. 2021a} and \citet{Yang 2021b}. Therefore, we do not introduce them here. In this paper, we focus on the large scale propagation of the winds driven by line force. The purpose of carrying out small scale simulations is to provide inner boundary conditions for the large scale simulations.

The general results are summarized in Table 1. The properties of the winds in Table 1 are obtained by time-averaging the values at the outer radial boundary on a time interval of $1.0T_{ob}-2.0T_{ob}$, with $T_{ob}$ being the orbital time at the radial outer boundary. Throughout the entire article, we give the time-averaged values on the range of $1.0T_{ob}-2.0T_{ob}$. From Table 1, we can see that for most of the models (except model RHD63), the mass flux of the winds measured at the outer radial boundary of the small scale simulations roughly equals to that measured at the outer boundary of the large scale simulations. This indicates that in these models, the winds generated quite close the black hole can move to significantly large radii ($>10^6 r_s$). The behaviors of the winds at large scale in these models are quite similar. We just take model RHD86 as a fiducial model to illustrate the large scale dynamics of the winds in these models. It is also clear that for the model RHD63, the winds generated close to the black hole can not move to large radii. In the following, we will also discuss why in model RHD63, the large scale dynamic of the winds is different.

\subsection{Fiducial Model RHD86}

Figure \ref{Fig:dv86} shows the time-averaged density (color) and velocity (arrows) for model RHD86. Left panel is for small scale simulations and right panel is for large scale simulations. As found by \citet{Proga et al. 2000} and \citet{Nomura et al. 2016}, the line force driven winds mainly come from the region inside $200r_s$ (left panel). The winds are located in a angular region of $\theta > 45^\circ$. When the winds move to large radii ($> 1500r_s$), their angular locations do not change.

It is clear that the radial velocity of the winds at radius $10^5r_s$ is much larger than that of the winds at $1500r_s$. This indicates that the winds are accelerated when they move in the region from $1500r_s$ to $10^5r_s$. In order to show this point more clearly, we plot Figure \ref{Fig:Bernoulli}. In this figure, we plot the time-averaged radial dependence of kinetic energy $1/2(\upsilon/c)^2$ and gravitational energy $-GM/(r-r_s)c^2$ along $\theta=63^\circ$. At this angle, the winds move almost radially. Therefore, we can take Figure \ref{Fig:Bernoulli} as the results along a streamline of the winds. The kinetic energy of the winds is significantly larger than their gravitational energy. Thus, the winds are unbound to the black hole potential. It is clear that the kinetic energy of the winds keeps increasing from $1500r_s$ to $10^5r_s$. The accelerating force is radiation line force. Outside $10^5r_s$, the kinetic energy of the winds does not change with the increasing of radius. Therefore, the acceleration line force vanishes outside $10^5r_s$. The line force is $\textbf{M} F_e$, with $F_e$ being the radiation force due to electron scattering and $\textbf{M}$ being the line force multiplier. The line force multiplier is a function of both ionization parameter and temperature of gas (\citealt{Proga 2007}).

\begin{figure*}
\begin{center}
\includegraphics[width=0.45\textwidth]{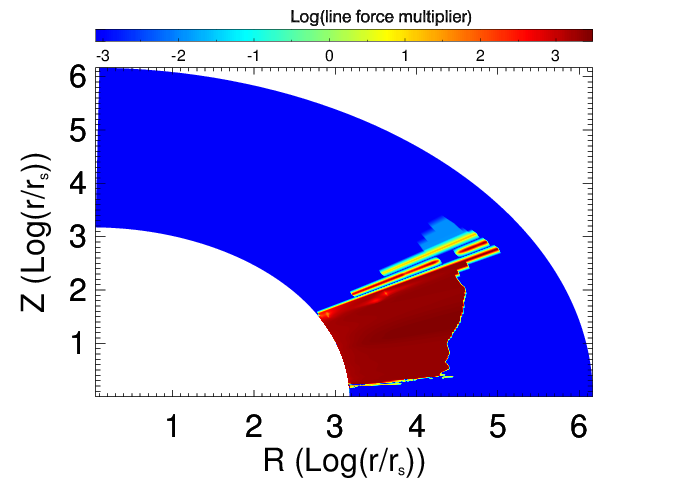}
\includegraphics[width=0.45\textwidth]{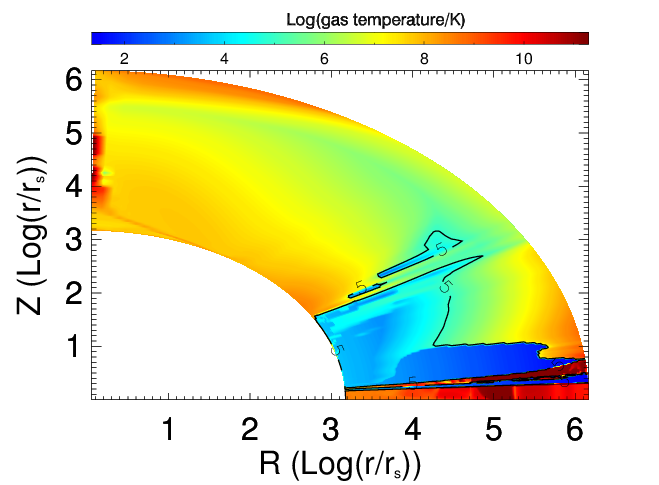}
\caption{Time-averaged logarithm of line force multiplier (left panel) and gas temperature (right panel) for model RHD86. The black line in the right panel shows the contour of temperature of $10^5$K. \label{Fig:mul}}
\end{center}
\end{figure*}

In Figure \ref{Fig:mul}, we plot the time-averaged logarithm of line force multiplier (left panel) of model RHD86. It is clear that inside $10^5r_s$, in the winds region ($\theta>60^\circ$), the line force multiplier is several orders of magnitude higher than $1$. Therefore, the winds are accelerated when they move from small to large radii inside $10^5r_s$. Outside $\sim 10^5r_s$, the line force multiplier is negligibly small. In order to find out the reason why outside $10^5r_s$ in the winds region ($\theta>60^\circ$), the line force multiplier is negligible, we plot the right panel of Figure \ref{Fig:mul}. This panel plots the logarithm of gas temperature. As shown in this panel, outside $10^5r_s$ in the winds region ($\theta>60^\circ$), the gas temperature is higher than $10^5$K. Thus, the line force is negligible outside $10^5r_s$, and the winds move outwards with constant velocity. We point out that the high temperature of gas is due to photoionization heating by the X-rays. We check the simulation data and find that the high temperature region and high ionization parameter region are almost overlapped. In other words, X-ray photoionization is the primary reason for the small force multiplier.

\begin{figure*}
\begin{center}
\includegraphics[width=0.3\textwidth]{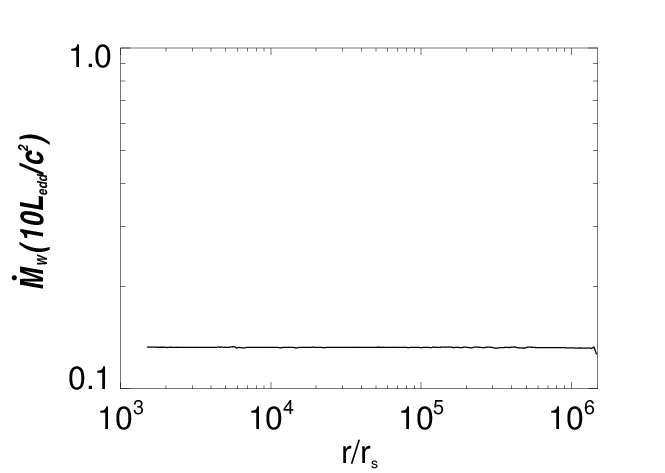}
\includegraphics[width=0.3\textwidth]{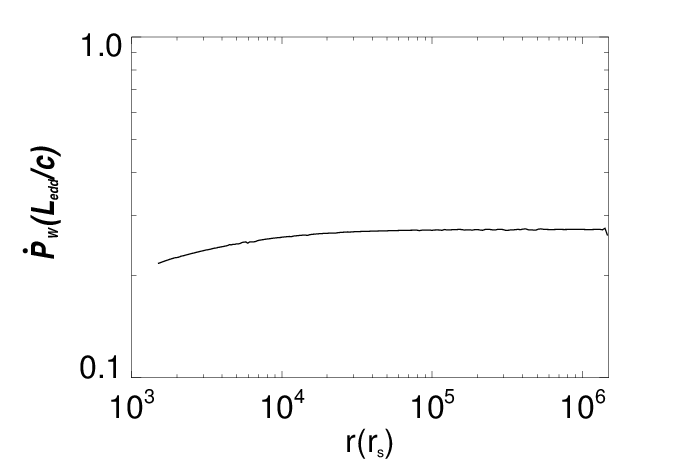}
\includegraphics[width=0.3\textwidth]{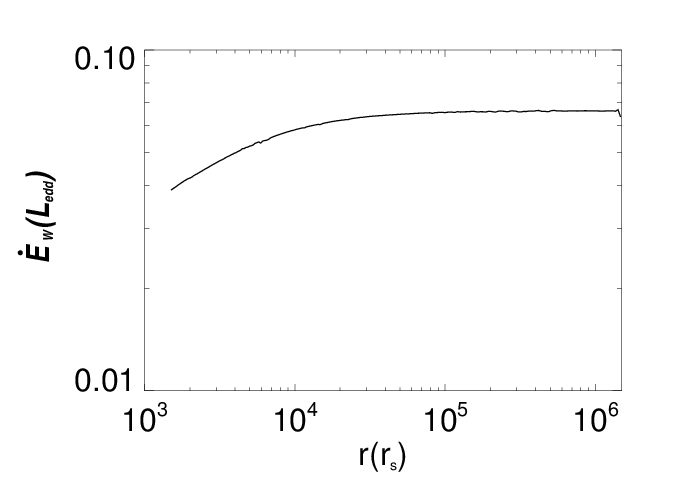}
\caption{Time-averaged mass flux (left panel), momentum flux (middle panel) and kinetic power (right panel) of winds for model RHD86.  \label{Fig:mdotpdotedot}}
\end{center}
\end{figure*}

Figure \ref{Fig:mdotpdotedot} shows the time-averaged radial profiles of the winds mass flux (left panel), momentum flux (middle panel) and kinetic energy flux (right panel) for model RHD86. The mass flux, momentum flux and kinetic energy flux are calculated as follows
\begin{equation}
\dot M_{\rm w}(r)=4\pi r^{2} \int_{\rm 0^\circ}^{\rm 89^\circ} \rho \max(\upsilon_r,0) \sin\theta d\theta ,
\label{mdot}
\end{equation}
\begin{equation}
\dot P_{\rm w}(r)=4\pi r^{2} \int_{\rm 0^\circ}^{\rm 89^\circ} \rho \max(\upsilon_{r}^{2},0) \sin\theta d\theta ,
\label{pdot}
\end{equation}
\begin{equation}
\dot E_{\rm w}(r)=4\pi r^{2} \int_{\rm 0^\circ}^{\rm 89^\circ} \frac{1}{2}\rho \max(\upsilon_{r}^{3},0) \sin\theta d\theta .
\label{edot}
\end{equation}
In these calculations, the integrations stop at $\theta=89^\circ$ to exclude the contribution of the accretion disk at the midplane.

The mass flux is a constant with radius, which indicates that all of the winds generated at small scale (inside $200r_s$) can escape to large radii. They do not stop at the outer radial boundary of our simulations. The winds can move out of the AGNs to interact with the interstellar medium of their host galaxies. Both the momentum flux and kinetic power of the winds keep increase from $1500r_s$ to $10^5r_s$ due to the acceleration by line force. Outside $10^5r_s$, the acceleration vanishes and the momentum flux and kinetic power do not change with the increase of radius. The kinetic power of the winds is significantly higher than $0.5\%$ of the Eddington luminosity, therefore, the winds can give sufficient feedback to their host galaxies (e.g., \citealt{Di Matteo et al. 2005}; \citealt{Hopkins and Elvis 2010}; \citealt{Ostriker et al. 2010}).

\begin{figure*}
\begin{center}
\includegraphics[width=0.45\textwidth]{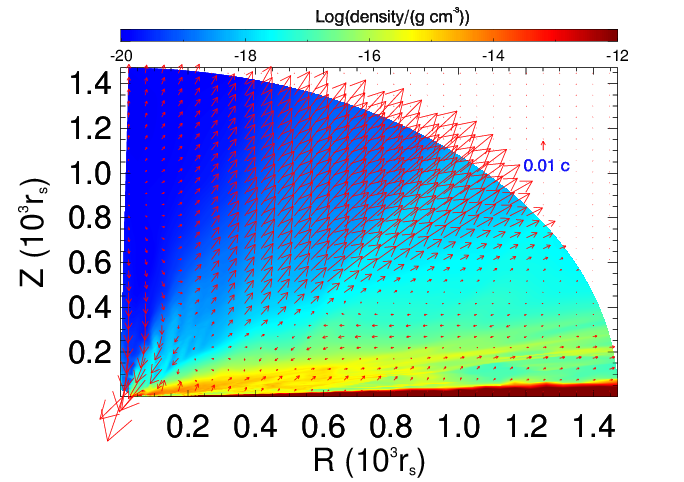}
\includegraphics[width=0.45\textwidth]{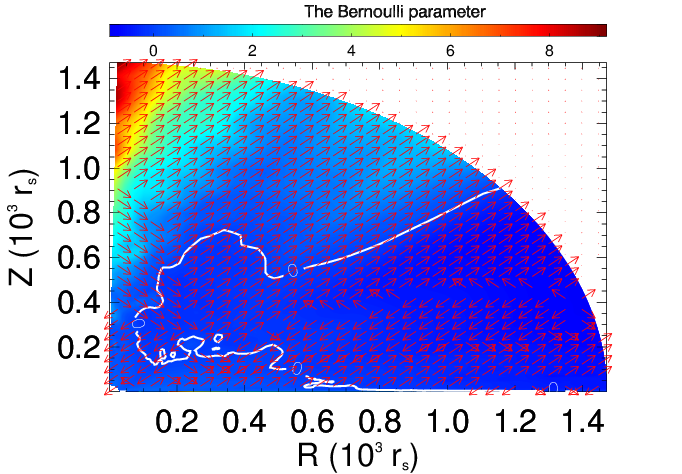}
\includegraphics[width=0.45\textwidth]{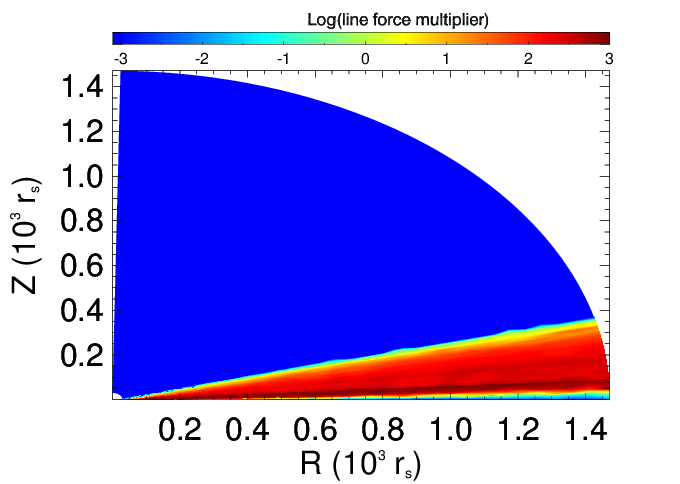}
\includegraphics[width=0.45\textwidth]{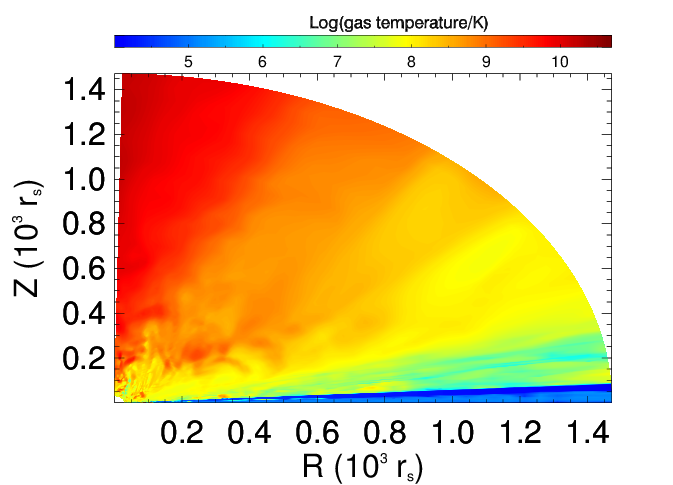}
\includegraphics[width=0.45\textwidth]{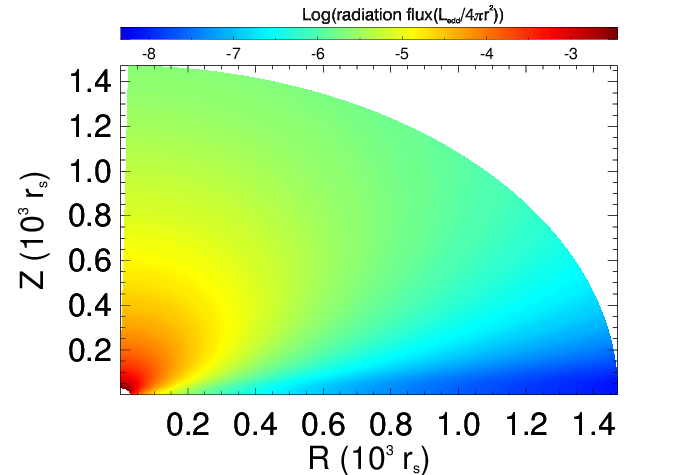}
\includegraphics[width=0.45\textwidth]{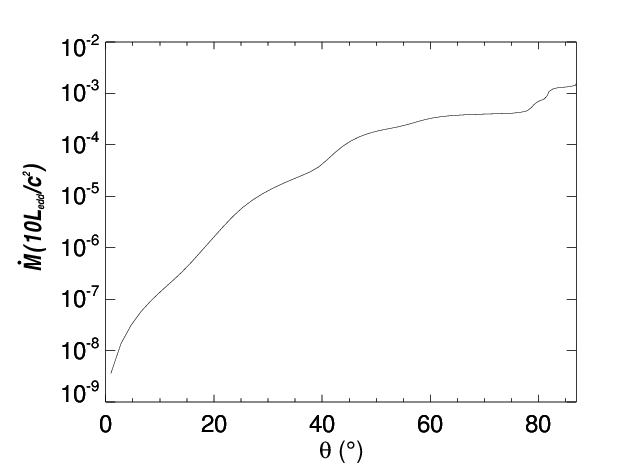}
\caption{Properties for model RHD63. Top left: time-averaged density (color) and velocity (arrows). Top right: time-averaged the Bernoulli parameter (color) and unit vector of poloidal velocity (arrows). The white solid line shows that the contour of the value of the Bernoulli parameter equals 0. Middle left: time-averaged logarithm of line force multiplier. Middle right: time-averaged logarithm of gas temperature. Bottom left: logarithm of the radiation flux from the accretion disk in units of $L_{edd}/4\pi r^2$. Bottom right: time-averaged the angular dependence of the mass flux of winds measured at $1500r_s$.  \label{Fig:model63-1}}
\end{center}
\end{figure*}

\begin{figure*}
\begin{center}
\includegraphics[width=0.45\textwidth]{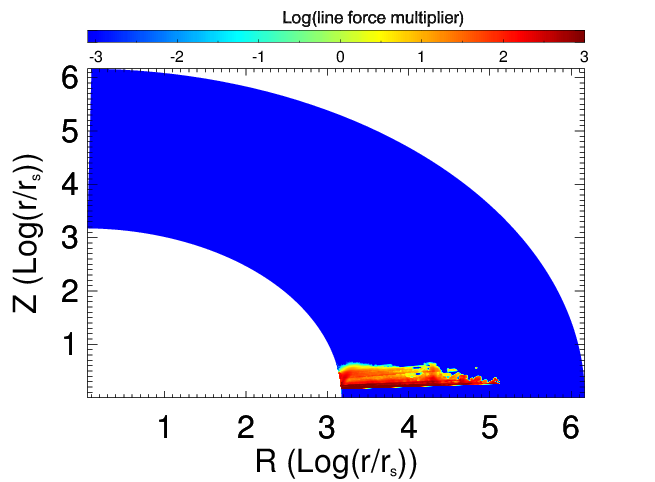}
\includegraphics[width=0.45\textwidth]{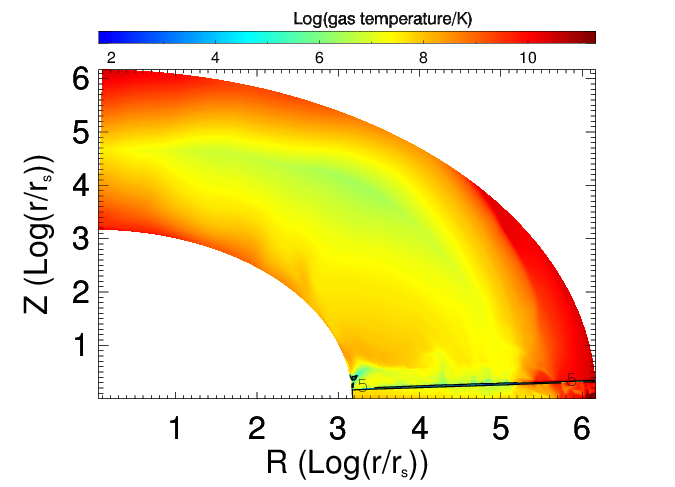}
\caption{Time-averaged large scale simulations (from $1500r_s$ to $1.5\times10^6r_s$) results of logarithm of line force multiplier (left panel) and gas temperature (right panel) for model RHD63. The black line in the right panel shows the contour of temperature of $10^5$K. \label{Fig:mul63-large}}
\end{center}
\end{figure*}

\subsection{Varying Black Hole Mass and Bolometric Luminosity of Accretion Disk}
From table 1, we can see that the properties of the winds (mass, momentum and kinetic energy fluxes) depend on the Eddington ratio  of radiation flux and the black hole mass. Models with higher Eddington ratio of radiation flux have stronger winds. In the models with $\varepsilon = 0.6$, the mass flux of the winds exceeds $10\%$ of the Eddington accretion rate. Also, the kinetic energy flux of the winds measured exceeds $1\%$ of the Eddington luminosity. In the models with $\varepsilon = 0.3$, the mass, momentum and kinetic energy fluxes (in unit of Eddington values) are smaller. This is easy to understand that stronger radiation flux can drive stronger winds. The dependence of the winds properties on black hole mass is generally that the smaller the black hole mass is, the weaker the winds will be. Giustini et al. (2019) also find that the smaller black hole mass will produce weaker winds. The reasons are as follows. The temperature of the accretion disk $\propto I_D^{1/4}$, and the accretion disk radiation intensity depends on black hole mass (see Equation 4). For a larger $M_{BH} \geq 10^8 M_\odot$, the peak disk temperature is around optical/UV; while for a smaller $M_{BH} \leq 10^8 M_\odot$, it moves to far UV/soft X-ray regime. Therefore, smaller black hole will produce relatively weaker UV flux and stronger X-ray flux. Weaker UV flux will exert weaker line force, and stronger X-ray flux will over-ionize gas to a level where line force is no longer effective.

For most of the models (except model RHD63), the evolutions of the dynamics of the winds at large scale are quite similar, which have been explained by using the fiducial model RHD86.

In model RHD63, the mass flux of the winds (in unit of Eddington mass flux) is one or two orders of magnitude smaller than that of the winds in other models. Also, from table 1, it is clear that the mass flux of the winds measured at $1.5\times 10^6r_s$ is almost one order of magnitude smaller than that measured at $1500r_s$. About $90\%$ of the winds generated close to the black hole can not move to large radii. In the following, we give the reasons.

We plot some time-averaged properties of model RHD63 in Figure \ref{Fig:model63-1}. Comparing the upper left panel of this figure to the left panel of Figure \ref{Fig:dv86}, we can see that the velocity of the winds in model RHD63 is almost one order of magnitude smaller than that of the winds in model RHD86. The velocity of the winds in the region $\theta>45^\circ$ is even smaller. The usually adopted criterion to judge whether winds can escape is the Bernoulli parameter, which is defined as follows
\begin{equation}
Be=\frac{1}{2}(\upsilon_r^2+\upsilon_\theta^2+\upsilon_\phi^2)+\frac{\gamma P}{(\gamma-1)\rho}-\frac{GM_{BH}}{r-r_c}.
\end{equation}

The upper right panel of of Figure \ref{Fig:model63-1} shows the Bernoulli parameter in unit of gravitational energy. It can been seen that in the region $\theta>45^\circ$, the winds have a negative Bernoulli parameter. If there is no further acceleration, the winds can not escape and move to large radii. The negative Bernoulli parameter is due to the fact that the velocity (or kinetic energy) of the winds is too small. The middle left panel of Figure \ref{Fig:model63-1} shows the line force multiplier, it is obvious that the line force multiplier in the region $0^\circ-77^\circ$ is almost 0. The reason is that the gas temperature in this region (middle right panel) significantly higher than $10^5$K. The bottom left panel shows the logarithm of the radiation flux in unit of $L_{edd}/4\pi r^2$ from the accretion disk. The radiation flux in unit of $L_{edd}/4\pi r^2$ equals the ratio of the radiation force due to electron scattering to the gravity. In the region $\theta>77^\circ$, the line force multiplier can be much large. However, in this region, the radiation flux (in unit of $L_{edd}/4\pi r^2$) is smaller than $10^{-5}$. The radiation force due to electron scattering is more than 5 orders of magnitude smaller than gravity. Even though, in this region, we have the multiplier of several hundreds, the line force is still significantly smaller than gravity. Therefore, the winds in the region $\theta>45^\circ$ can not be effective accelerated by radiation pressure either due to the negligible line force multiplier (in the region $45^\circ<\theta<77^\circ$) or due to the negligible radiation flux (in the region $\theta>77^\circ$). The winds in the region $\theta>45^\circ$ are gravitationally bound to the central black hole. From the velocity vector of top right panel in Figure \ref{Fig:model63-1}, the winds in this region are much like the outwards moving part of some turbulent eddies.

The bottom right panel of Figure \ref{Fig:model63-1} shows the $\theta$ direction profile of the time-averaged mass flux of the winds at $1500r_s$. It is clear that most of the mass flux is in the region $\theta>45^\circ$. As explained above, the winds in the region ($\theta>45^\circ$) are gravitationally bound to the black hole. If the winds in this region can not be accelerated by radiation pressure in the region $r>1500r_s$, they will not escape and fall back. Figure \ref{Fig:mul63-large} shows the time-averaged logarithm of line force multiplier (left panel) and gas temperature (right panel) in the region $r>1500r_s$ for model RHD63. It is clear that at large radii ($r>1500r_s$), the line force multiplier is only important in a very limited region close to the midplane. In other region, the temperature of gas significantly exceeds $10^5$K, so that the force multiplier is negligible. In the region where line force multiplier is large, the radiation flux is small, and the radiation pressure is not important. Therefore, outside $1500r_s$, the winds can not be continuously accelerated. As listed in Table 1, in model RHD63, about $90\%$ of the winds generated close to the black hole can not move to large radii. The kinetic power of the winds in this model is several order of magnitude lower than $0.5\%$ of the Eddington luminosity. The winds can not provide sufficient feedback to their host galaxies.

\subsection{Comparison to Observations}
\begin{figure*}
\begin{center}
\includegraphics[width=0.45\textwidth]{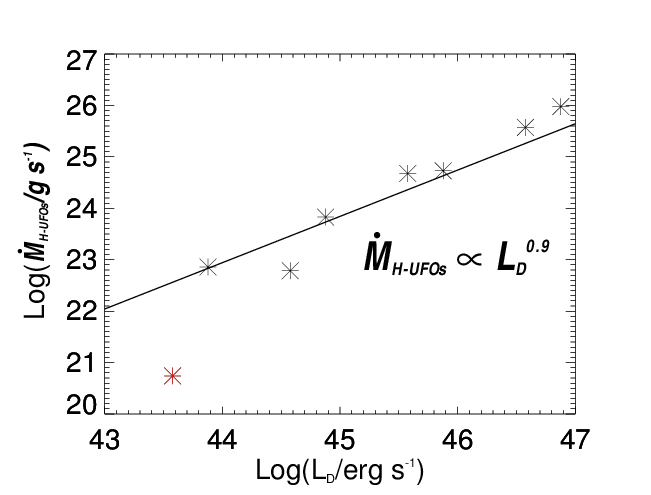}
\includegraphics[width=0.45\textwidth]{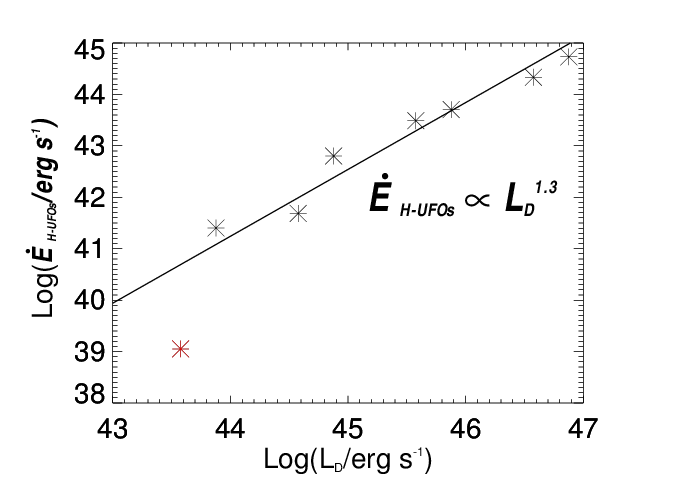}
\caption{Left panel: time-averaged mass flux of H-UFOs (measured at the outer radial boundary of the small scale simulation) versus accretion disk luminosity. Right panel: time-averaged kinetic energy flux of H-UFOs (measured at the outer radial boundary of the small scale simulation) versus accretion disk luminosity. The asterisks correspond to our simulation results. The solid lines correspond to our power law fitting. The red asterisks correspond to model RHD63, which has very weak winds. Also, the results in model RHD63 deviate from the fitting power-law significantly. \label{Fig:HUFO}}
\end{center}
\end{figure*}

\begin{figure*}
\begin{center}
\begin{minipage}[!htbp]{0.45\linewidth}
\includegraphics[width=1.0\textwidth]{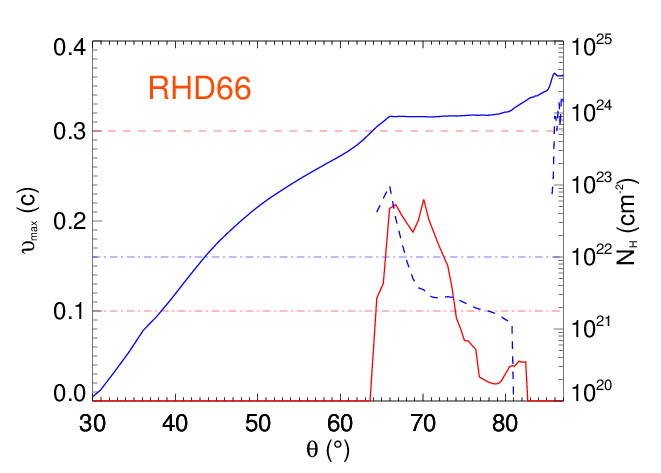}

\includegraphics[width=1.0\textwidth]{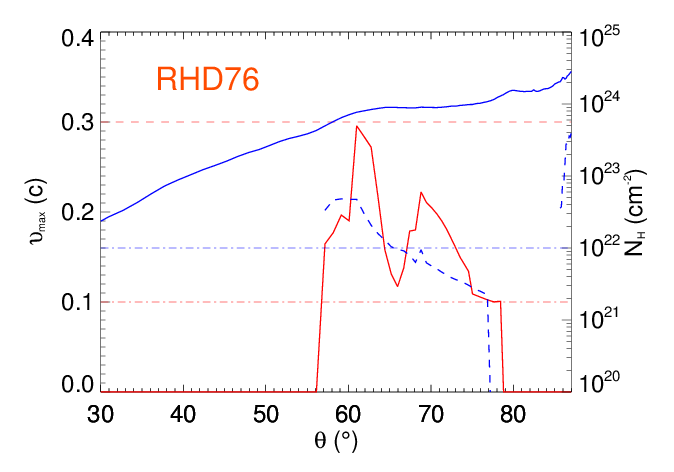}

\includegraphics[width=1.0\textwidth]{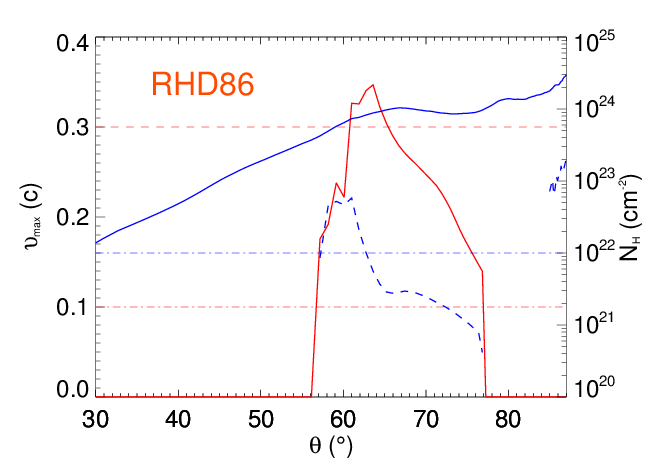}

\includegraphics[width=1.0\textwidth]{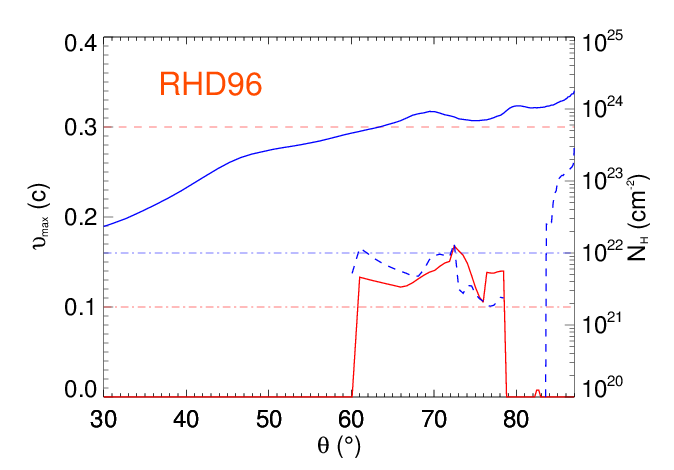}
\end{minipage}
\begin{minipage}[!htbp]{0.45\linewidth}
\includegraphics[width=1.0\textwidth]{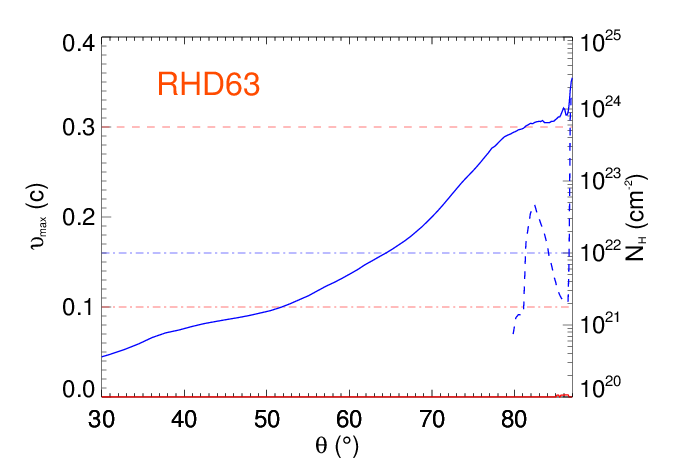}

\includegraphics[width=1.0\textwidth]{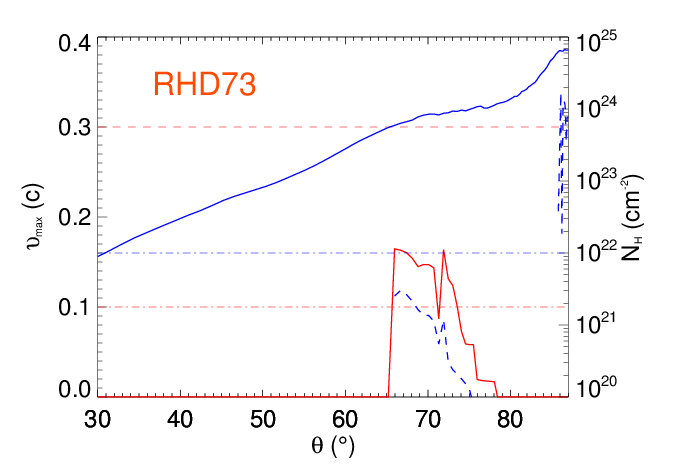}

\includegraphics[width=1.0\textwidth]{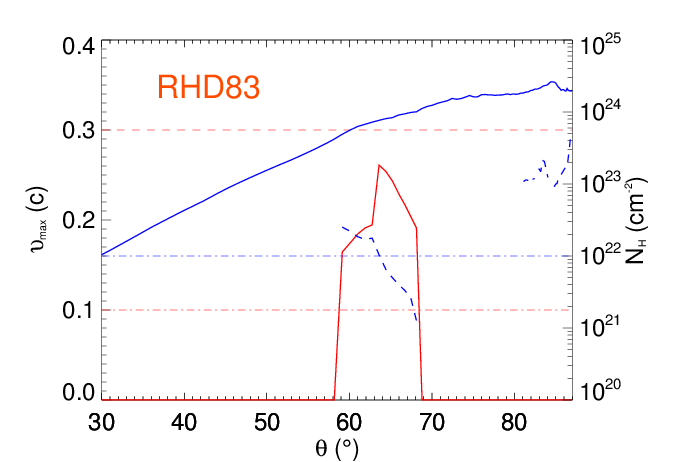}

\includegraphics[width=1.0\textwidth]{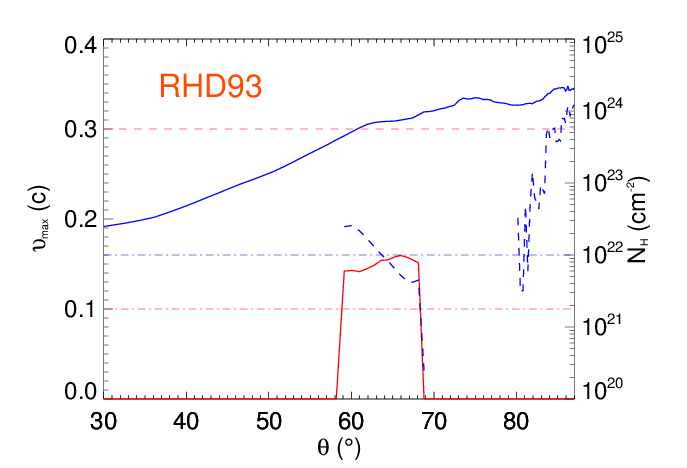}
\end{minipage}
\caption{The $\theta$ dependence of time-averaged column density (solid blue line), column density of gas with $\log \xi /(\text{erg cm s}^{-1})<2$ (dashed blue line), maximum velocity of winds with $\log \xi /(\text{erg cm s}^{-1})<2$ (solid red line) at the outer boundary of the large scale simulations. The red dotted-dashed line, the red dashed line and the blue dotted-dashed line show the lines of $\upsilon_{max}=0.1\text{c}$, $\upsilon_{max}=0.3\text{c}$ and $N_{H}=10^{22}\text{cm}^{-2}$ , respectively. The ordinate on the left-hand side shows the velocity. The ordinate on the right-hand side shows the column density.  \label{Fig:efuo}}
\end{center}
\end{figure*}
We find that the ionization parameter of the winds changes with time. The reasons are as follows. All of the winds materials come from the midplane accretion disk. In order to form the winds, the gas should be first puffed up from the accretion disk by line force. If there is blocking gas located between the puffed-up gas and the central black hole, the X-rays will be blocked and the puffed-up gas will have lower ionization parameter. In this case, the puffed-up gas will be accelerated to high velocity. When the blocking gas disappears, the high velocity winds will be exposed to the central X-rays. The high velocity winds become highly ionized. Because of the turbulent nature of the flow, the blocking gas is not hard to form. The blocking gas can both appear and disappear with time. Therefore, the ionization parameter changes with time.

Ultra-fast outflows (UFOs) absorbers have been found in hard X-ray bands in $\sim 40$ percents of nearby AGNs (\citealt{Pounds et al. 2003}; \citealt{Tombesi et al. 2010,Tombesi et al. 2011,Tombesi et al. 2012}; \citealt{Gofford et al. 2013,Gofford et al. 2015}; \citealt{Laha et al. 2021}). We define the UFOs found in hard X-ray bands as H-UFOs. Their properties are as follows. The velocity is in the range of  $\upsilon\sim 0.03 -0.3 \text{c}$. The ionization parameter is in the range of $\xi\sim 10^{3}-10^{6} \text{erg cm s}^{-1}$. The column density is in the range of $10^{22} \text{cm}^{-2}\leq N_{H}\leq 10^{24} \text{cm}^{-2}$. The launching locations of H-UFOs are close to or smaller than hundreds Schwarzschild radius. It is found that the H-UFOs may be driven from accretion disk by line force (\citealt{Nomura et al. 2016}; \citealt{Nomura and Ohsuga 2017}; \citealt{Yang et al. 2021a}). In our small scale simulations, we do find that H-UFOs are present. The H-UFOs correspond to the winds which at the observational epoch have high ionization parameter. \citet{Gofford et al. 2015} analyze the observational data of H-UFOs in a sample. They find the mass flux of H-UFOs $\dot M_{H-UFOs} \propto L_D^{0.9}$ and the kinetic power of H-UFOs $\dot E_{H-UFOs} \propto L_D^{1.5}$. We also calculate the mass flux and kinetic power of the H-UFOs based on our small scale simulations. At each snapshot, we first find out winds which satisfy the conditions of H-UFOs. Then we calculate their mass flux and kinetic power. Finally, we time-average the mass flux and the kinetic power. The results are shown in Figure \ref{Fig:HUFO}. In this figure, the asterisks correspond to our simulations. The solid line is the fitting power law function. It is clear that except model RHD63, all other models can be well described by the fitting power-law function. The power law indexes are almost same as those given by observations (\citealt{Gofford et al. 2015}). Moreover, \citet{Nomura and Ohsuga 2017} find the similar results. Therefore, the H-UFOs are quite probably driven by line force.

In addition to H-UFOs found in hard X-ray bands, ultra-fast outflows have also been found in the soft X-ray bands (e.g., \citealt{Gupta et al. 2013,Gupta et al. 2015}; \citealt{Longinotti et al. 2015}; \citealt{Pounds et al. 2016}; \citealt{Reeves et al. 2016}; \citealt{Serafinelli et al. 2019}). We define the ultra-fast outflows in soft X-ray bands as S-UFOs. The S-UFOs have velocity $\upsilon \sim 0.1-0.2 c$, ionization parameter $\log \xi \sim <2 \,\text{erg cm s}^{-1}$ and column density $ N_H \sim 10^{20} \text{cm}^{-2} - 10^{22} \text{cm}^{-2}$. The locations of the S-UFOs are quite hard to be constrained. The S-UFOs may be located around (\citealt{Reeves et al. 2016}) or outside (\citealt{Serafinelli et al. 2019}) parsec scale. \citet{Serafinelli et al. 2019} propose that the S-UFOs are generated by the interaction between H-UFOs and the interstellar medium (ISM). The S-UFOs are the entrained ISM by the H-UFOs.

In this paper, we investigate another possibility that S-UFOs are actually the winds driven by line force. As mentioned above, in the small scale simulations, we find that the ionization parameter of the winds can change with time. The S-UFOs may correspond to the winds which at the observational epoch have low ionization parameter. In order to test this idea, we plot Figure \ref{Fig:efuo}. This figure shows the view angle ($\theta$) dependence of column density and velocity of the winds. When plotting this figure, the column density is obtained by integration from $30-10^6r_s$. It can been seen that in most models (except RHD63), we can observe low ionization $\xi < 10^2 \text{erg cm s}^{-1}$ winds. The low ionization winds have velocity $\upsilon \sim 0.1c-0.3 c$ and column density $N_H \sim 10^{20} \text{cm}^{-2} - 10^{22} \text{cm}^{-2}$. The properties of the low ionization winds are in the right property scope of the observed S-UFOs. By observations, one can obtain the ionization parameter, velocity and column density of winds. However, it is quite hard to estimate the mass flux of winds observationally. The reason is that in order to quantify the mass flux, one need to know the covering and volume filling factors of winds. The covering factor is usually set to be equal to the ratio of the sources which show winds to the total sources of the sample. The volume filling factor can not be well constrained. Due to the difficulties, many literatures of observations do not give the value of mass flux of winds. Therefore, we can not directly compare the mass flux of the S-UFOs found in this paper to observations. We can just conclude that there is possibility that the observed S-UFOs are just the low ionization accretion disk winds.

Slowly moving ($100-2000$ km s$^{-1}$) warm absorber outflows are common phenomena of AGNs (e.g., \citealt{Canizares and Kruper 1984}; \citealt{Sako et al. 2001}; \citealt{Blustin et al. 2004,Blustin et al. 2005}; \citealt{Reeves et al. 2004}; \citealt{McKernan et al. 2007}; \citealt{Laha et al. 2016}). Recently, by performing simulations, we find that when H-UFOs move to parsec scale, they can collide with ISM and transfer their momentum to ISM. The properties of the ISM after colliding with H-UFOs are quite similar as those of warm absorbers. We conclude that the warm absorber outflows may be the accelerated ISM by the H-UFOs (\citealt{Bu and Yang 2021}).

\section{Discussions}
In this paper, we time-average the simulation results at the outer boundary of the small scale simulations. As mentioned above, in the small scale simulations, both high and low ionization winds are present. The high ionization winds usually have low density; The low ionization winds usually have high density. At a snapshot, at a given location, there is the high ionization winds. At another snapshot, the low ionization winds may occupy this location. Therefore, if we do the time-average, the information of the high ionization winds can be removed due to its low density. Therefore, actually, in the inner radial boundary of our large scale simulations, we mainly inject the low ionization winds. In other words, in the large scale simulations, we mainly study the dynamics of the low ionization winds. The large scale dynamics of the high ionization winds have not been studied in this paper. In order to properly study the large scale dynamics of both low and high ionization winds simultaneously, one should carry out simulations with inner radial boundary located at close to the black hole and outer radial boundary located at far beyond parsec scale. However, such simulations are too expensive to be carried out. Therefore, we must keep in mind that based on this paper, we can not have any conclusion about the large scale dynamics of high ionization winds from the small scale accretion disk.

In this paper, as down by \citet{Proga et al. 2000} and \citet{Nomura and Ohsuga 2017}, we set $\sigma_{X}=\sigma_{e}$ for $\xi\geq10^{5}\text{erg cm s}^{-1}$ and $\sigma_{X}=100\sigma_{e}$ for $\xi<10^{5}\text{erg cm s}^{-1}$ . In \citet{Proga and Kallman 2004}, they set $\sigma_{X}=\sigma_{e}$ for entire range of $\xi$. \citet{Nomura et al. 2020} compare the results of the two methods. They find that both of the mass flux and the velocity of winds differ roughly by a factor of 2. However, we note that both of the two methods are arbitrary. In this paper, the scattered and reprocessed photons are neglected. In reality, the scattered and the reprocessed photons should also affect the properties of winds. For example, \citet{Higginbottom et al. 2014} find that the scattered photons can also ionize gas, which can make the gas too highly ionized to be driven by line force. However, we note that radiative transfer and hydrodynamics are decoupled in \citet{Higginbottom et al. 2014}. Therefore, in future, it is necessary to perform hydrodynamic simulations with radiative transfer to study the effects of the scattered and reprocessed photons.

In addition, we only consider the large scale dynamics of accretion disk winds driven by line force. In reality, the magnetic field is always be present and responsible for angular momentum transfer of the accretion disk. The properties of the winds in the presence of magnetic field should be different (\citealt{Cui et al. 2020b}; \citealt{Yang et al. 2021a}; \citealt{Yang 2021b}). In future, we plan to study the large-scale dynamics of winds driven by line force and Lorentz force.

\section{Summary}
We carry out two-dimensional simulations to study the large-scale dynamics of accretion disk winds driven by line force. The disk winds are launched from inside hundreds of Schwarzschild. We study whether the winds can move to large scale and what the properties of the winds will be at large scale. By large scale, we mean around or larger than parsec scale. It is currently impossible to perform simulations with the inner boundary close to the black hole and outer boundary around parsec scale. Therefore, we divide the space scale into two parts. We perform both small scale and large scale simulations. For the small scale simulations, our computational domain is $30r_s \leq r \leq 1500r_s$. The winds are produced in the small scale simulations. For our large scale simulations, the computational domain is $1500r_s \leq r \leq 1.5\times 10^6r_s$. In the large scale simulations, we inject winds at the inner radial boundary ($1500r_s$). The properties of the injected winds are obtained by the small scale simulations. By using the large scale simulations, we can study the large scale dynamics of winds.

We find that both the black hole mass and the accretion disk luminosity can affect the properties of the winds driven by line force. In our models, the black hole mass is in the range $10^6M_\odot-10^9M_\odot$. The accretion disk luminosity is set to 0.3 ($\varepsilon=0.3$) and 0.6 ($\varepsilon=0.6$) times the Eddington luminosity.

For the case $\varepsilon=0.6$, we find that independent of the black hole mass, the winds have mass flux exceeding $10\%$ the Eddington accretion rate and kinetic energy flux exceeding $1\%$ the Eddington luminosity. We also find that in this case, the accretion disk winds can go to outside $10^6r_s$. When arriving $10^6r_s$, the kinetic energy flux of the winds slightly increases compared to that at the launching point. The increase of the kinetic energy flux is due to the line force acceleration when the winds move outwards. The winds can have sufficient feedback to their host galaxies due to their high kinetic energy flux (e.g., \citealt{Di Matteo et al. 2005}; \citealt{Hopkins and Elvis 2010}).

For the case $\varepsilon=0.3$, the mass, momentum and kinetic energy fluxes of the winds are smaller than those in the case $\varepsilon=0.6$. We also find that for $\varepsilon=0.3$, the strength of the winds decreases with the decrease of black hole mass. For example, if we set black hole mass to be $10^8M_\odot$ and $10^9M_\odot$, the mass flux is roughly $6\%$ of the Eddington accretion rate; the kinetic energy flux is roughly $1\%$ the Eddington luminosity. For these two black hole masses, all the winds can move to the region outside $10^6r_s$. However, when the black hole mass is $10^7M_\odot$, the winds mass flux is significantly decreased to $1\%$ of the Eddington accretion rate, and the kinetic energy flux is decreased to $0.2\%$ of the Eddington luminosity. For this black hole mass, we also find that the winds can move to the region outside $10^6r_s$. If we further decrease the black hole mass to $10^6M_\odot$, the mass flux of the winds is decreased to $1.5 \times 10^{-3}$ Eddington accretion rate, and the kinetic energy is almost 6 orders of magnitude smaller than the Eddington luminosity. Also, with black hole mass $10^6M_\odot$, about $90\%$ of the winds can not move to large radii. Therefore, for the case $\varepsilon=0.3$, when black hole mass is $\leq 10^7M_\odot$, the winds can not provide sufficient feedback to their host galaxies.

The ionization parameter of the winds changes with time. The UFOs observed in hard X-ray band may correspond to winds which at the observational epoch have high ionization parameter. We compare our simulation results to observations. We find that, for the UFOs found in hard X-ray bands, their observed dependence of kinetic power and mass flux on the accretion disk luminosity (\citealt{Gofford et al. 2015}) can be well produced by the line force driven winds model. This indicates that the UFOs found in hard X-ray band are very probably driven by line force. The UFOs observed in soft X-ray band may correspond to winds which at the observational epoch have low ionization parameter. We also compare our simulation results to the UFOs found in soft X-ray bands. It is found that the observed properties (column density, velocity, ionization parameter) of the UFOs in soft X-ray bands can be well explained by the line force driven winds model. Therefore, the observed UFOs in soft X-ray bands are probably the accretion disk winds driven by line force.

\section*{Acknowledgments}
D. Bu is supported by the Natural Science Foundation of China (grant 12173065). X. Yang is supported by the Natural Science Foundation of China (grant 11973018) and Chongqing Natural Science Foundation (grant cstc2019jcyj-msxmX0581). F. Yuan is supported in part by the Natural Science Foundation of China (grant 11633006, 12133008, 12192220, 12192223), and the Key Research Program of Frontier Sciences of CAS (No. QYZDJSSW-SYS008). W. Lin is supported by the Natural Science Foundation of China (grant 11973025).

\section*{DATA AVAILABILITY}
The data underlying this article will be shared on reasonable request to the corresponding author.

\appendix
\section{Comparison with previous works}
\subsection{Effects of numerical settings}
\begin{figure}
\begin{center}
\includegraphics[width=0.5\textwidth]{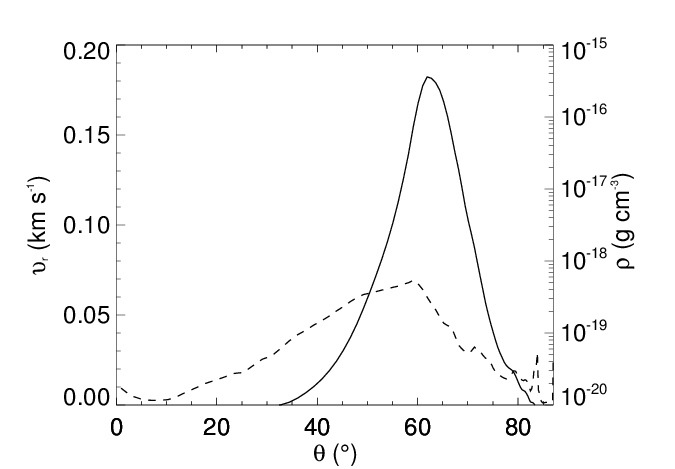}
\caption{Time-averaged angular profiles of the density and the radial velocity at $r=1500r_{s}$ for RHD83. The ordinate on the left-hand side of each panel refers to the solid line, while the ordinate on the right-hand side refers to the dotted line. \label{Fig1:cproga83}}
\end{center}
\end{figure}

\begin{figure*}
\begin{center}
\includegraphics[width=0.3\textwidth]{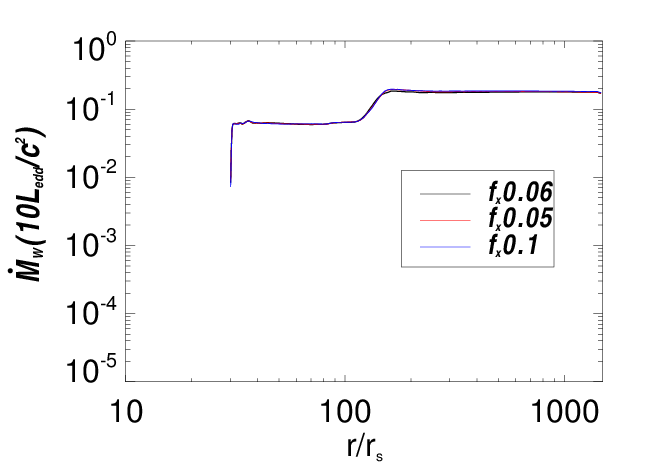}
\includegraphics[width=0.3\textwidth]{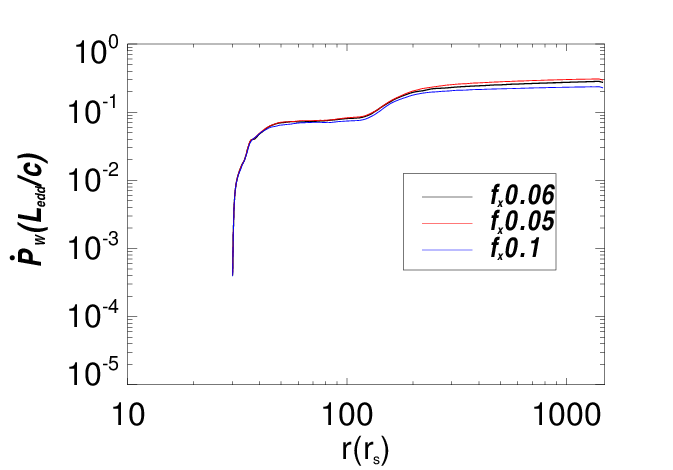}
\includegraphics[width=0.3\textwidth]{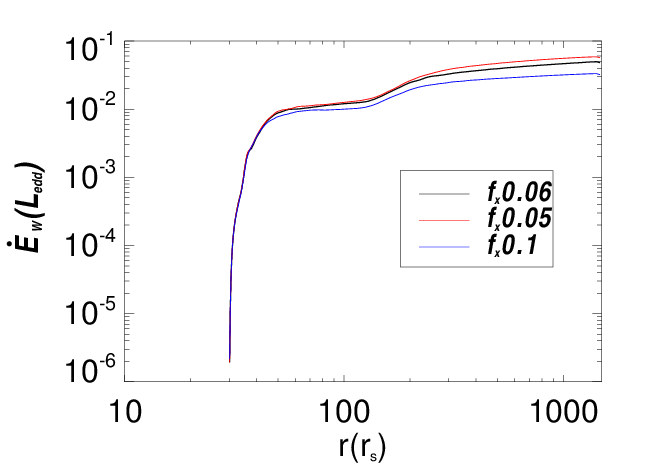}
\caption{Time-averaged mass flux (left panel), momentum flux (middle panel) and kinetic power (right panel) of winds for different values of $f_x$ at RHD86. The black lines correspond to $f_{x}=0.06$, which is calculated by referring to \citet{Yang et al. 2021a}, the red lines correspond to $f_{x}=0.05$, and the blue lines correspond to $f_{x}=0.1$. \label{Fig1:mdotpdotedot}}
\end{center}
\end{figure*}
Small scale simulations of line force driven winds have been studied intensively (e.g., \citealt{Proga et al. 2000}; \citealt{Proga and Kallman 2004}; \citealt{Nomura and Ohsuga 2017}; \citealt{Yang et al. 2021a}). There are three main differences of the settings of the previous simulation. The first one is the density of the accretion disk at the midplane. Some works assume that the density of the accretion disk is constant and does not vary with radius (\citealt{Proga et al. 2000}; \citealt{Proga and Kallman 2004}; \citealt{Yang et al. 2021a}). \citet{Nomura and Ohsuga 2017} set the density of the accretion disk at the midplane according to the accretion disk theory. We set the same accretion disk density as \citet{Nomura and Ohsuga 2017}(see Equation 5). The second one is the ratio of the X-ray luminosity to the accretion disk luminosity (the parameter $f_x$). Some works assume that the value of $f_x$ is fixed, which is independent from both the black hole mass and the accretion disk luminosity (e.g., \citealt{Proga et al. 2000}; \citealt{Proga and Kallman 2004};\citealt{Nomura and Ohsuga 2017}). \citet{Yang et al. 2021a} calculate the value of $f_x$ according to both black hole mass and accretion disk luminosity. The setting of $f_x$ in this paper is the same as \citet{Yang et al. 2021a}. The third one is the disk radiation contributing the line force. In \citet{Proga and Kallman 2004} and \citet{Nomura and Ohsuga 2017}, the disk radiation between 200-3200 {\AA} is contributing line force acceleration. In this paper (see also \citealt{Proga et al. 2000} and \citealt{Yang et al. 2021a}), it is assumed that radiation from high temperature region, in which the effective temperature is larger than 3000K, contributes the line force. We take model RHD86 as an example to illustrate that the two methods give roughly the similar radiation luminosity contributing the line force. The disk luminosity calculated by assuming all the radiation from the disk region with effective temperature higher than 3000K is $0.59 L_{edd}$ (model RHD86). When we keep other parameters (black hole mass, bolometric luminosity) same, we find that the disk luminosity between 200-3200 {\AA} is $0.45 L_{edd}$. Therefore, the difference of the disk luminosity calculated by the two methods differs by a factor smaller than 2. In their Appendix, \citet{Nomura et al. 2020} also find that the mass flux of winds roughly linearly depends on the disk radiation flux. Therefore, the slight difference of disk radiation obtained between the two methods will not significantly affect the wind properties.

We take model RHD83 as an example to compare our results to the results in previous works. In model RHD83, the black hole mass is $10^8M_\odot$; the value of $f_x \sim 0.09$; the accretion disk luminosity is $0.3 L_{edd}$. Also, in this model the density of the accretion disk is set according to the accretion disk theory. The settings of model RHD83 are quite similar to one of models in \citet{Nomura and Ohsuga 2017}. In \citet{Nomura and Ohsuga 2017}, they have a model for $M_{BH}=10^8M_\odot$, $f_x=0.1$, and $L_{D}=0.1L_{edd}$. Also in that model the density of the accretion disk is set according to the accretion disk theory. We call that model in \citet{Nomura and Ohsuga 2017} as model NO17comparison. In Figure \ref{Fig1:cproga83}, we plot the time-averaged angular profiles of density and velocity for model RHD83. We can compare this figure to the blue lines in Figure 5 in \citet{Nomura and Ohsuga 2017}. The blue lines in that figure are for model NO17comparison. The maximum velocity of winds in both model RHD83 and NO17comparison is $\sim 0.18c$. Also the maximum density in both of the models is several times of $10^{-19} $g/cm$^3$. The similarity of the results in both of the models indicates that the slight difference of numerical settings can not significantly affect the results.

We also study the effects of different values of $f_x$ on the results. In Figure \ref{Fig1:mdotpdotedot}, we plot the mass flux (left panel), momentum flux (middle panel) and kinetic energy flux (right panel) of winds. The black line is for model RHD86 and the value of $f_x=0.06$ in this model is calculated by Equation (6-8) in \citet{Yang et al. 2021a}. We also do two test simulations. In the two test simulations, we directly set the values of $f_x$ but with other parameters same as those in model RHD86. The blue ($f_x=0.1$) and red ($f_x=0.05$) lines correspond to the two test simulations. It can be seen from Figure \ref{Fig1:mdotpdotedot} that the slight change of the value of $f_x$ has negligible effects on the properties of winds.

\subsection{Force multiplier}
\begin{figure}
\begin{center}
\includegraphics[width=0.5\textwidth]{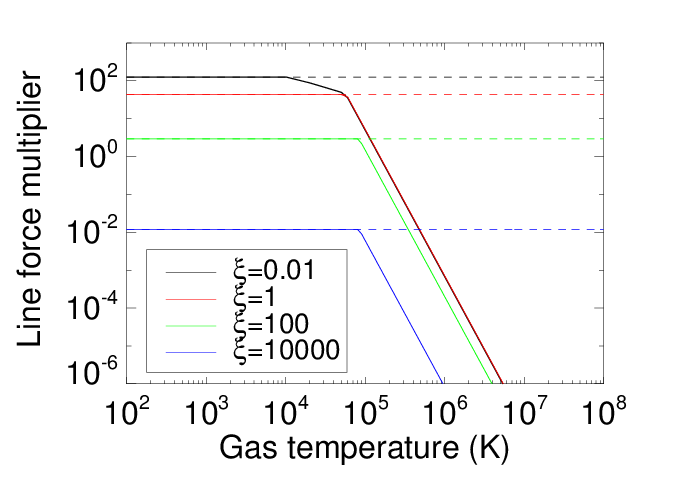}
 \caption{The force multiplier as functions of the gas temperature for model RHD86. The solid lines employ the Equation (17) of \citet{Proga 2007} (with temperature limitation), while the dot lines employ the Equation (12) of \citet{Proga 2007} (Without temperature limitation). The black solid, red solid, green solid and blue solid lines correspond to $\xi =0.01$, $\xi =1$, $\xi = 100$ and $\xi = 10000$, respectively. \label{Fig:multg}}
\end{center}
\end{figure}
The force multiplier depends both on the ionization parameter and gas temperature. The dependence of force multiplier on gas temperature is given by Equation (17) in \citet{Proga 2007}. According to that equation, the force multiplier will be negligible when gas has temperature above $10^5$K. In our present paper, we do not use the Equation (17) in \citet{Proga 2007} to constrain the force multiplier. Instead, we just assume that the force multiplier is zero when gas has temperature above $10^5$K. In order to justify our method, we plot Figure \ref{Fig:multg}. In this figure, we plot the variation of force multiplier with temperature. Because the force multiplier depends both on gas temperature and ionization parameter. When we calculate the temperature dependence, we need to fix the ionization parameter. The black solid, red solid, green solid and blue solid lines correspond to $\xi =0.01$, $\xi =1$, $\xi = 100$ and $\xi = 10000$, respectively. The dashed lines are for the cases without considering temperature dependence. From this figure, it is clear that below $10^5K$, the values of force multiplier in case considering the temperature dependence (Equation 17 in \citet{Proga 2007}) are roughly same as those in case without considering the temperature dependence. Therefore, below $10^5K$, it is not necessary to consider the temperature dependence of force multiplier. Above $10^5K$, it is reasonable to directly set force multiplier to be zero as in the present paper.

\end{document}